\documentstyle[12pt]{article}
\begin{document}
\newcommand{\vac}{{\rm (vac)}}
\begin{center}
{\Large{\bf Type-II parametric
down conversion in the Wigner-function formalism.
Entanglement and Bell's inequalities}}\\
Alberto Casado$^1$, Trevor W.~Marshall$^2$ and
Emilio Santos$^3$\\

$^1$ Escuela Superior de Ingenieros, Departamento de F=A1sica Aplicada,

Universidad de Sevilla,
Sevilla, Spain.

$^2$Departamento
de F\'{i}sica Moderna, Universidad de
Cantabria, 

39005 Santander, Spain.

$^3$Department of Mathematics, University of
Manchester, 

Manchester M13 9PL, U. K.

\vspace{0.5cm}

{\bf Abstract}
\end{center}
We continue the analysis of our
previous articles which were devoted to type-I
parametric down conversion, the
extension to type-II being straightforward.
We show that entanglement, in the Wigner
representation, is just a correlation
that involves both signals and vacuum fluctuations.
An analysis of the detection process
opens the way to a complete description of
parametric down conversion
in terms of pure Maxwell electromagnetic waves.

\newpage

\section{Introduction}
The theory of parametric down conversion (PDC) was treated, 
in the Wigner formalism,
in an earlier series of articles\cite{pdc1,pdc2,pdc3}. There we showed
that, provided one considers the zeropoint fluctuations of the
vacuum to be real, the description of radiation
is fully within Maxwell electromagnetic theory.
Effectively, because the Wigner function maintains its
positivity, we can say that
quantization is just the addition of a zeropoint radiation, and
there is no need for any further
quantization of the light field. In the present article we
show that the same result extends, without any difficulty,
from the type-I PDC case to the type-II situation.

There seems to be a widespread reluctance to accept the reality
of the vacuum fluctuations, in spite of the fact that they appear,
quite naturally, in the Wigner function of the vacuum state. We
remark that such fluctuations have been taken seriously,
within a certain school of thought, throughout
the entire history of the quantum theory, following the formulation
of Max Planck, originating in 1911\cite{mexicans}. Of course, it is true
that, integrated over all frequencies, they give us a vacuum with infinite
energy density; why then are all photographic plates not blackened
instantaneously? But all photodetectors, including even our own eyes,
are very selective, not only as regards the frequency, but even also the
wave vectors, of the light components they analyze. This is especially the
case with the detectors commonly used in PDC experiments. So, there is
a noise to subtract, but it is not infinite!

In our previous articles we indicated how the noise subtraction is
made, according to the Wigner formalism, and showed how this subtraction
is related to the standard calculating device, of normal ordering, used in
the Hilbert-space formalism. Here we extend this analysis, in an informal
manner, showing that, if we take into account the fact that all
detectors integrate the light intensity over a large time window, the
process of light detection, like that of light propagation, may also be
described
entirely in terms of real waves and positive probabilities.
We are then able to see that, in terms of a purely
wave description, the highly problematic concept
of ``entangled-photon" states of the field loses
all its mystery. Entangled photons are just
correlated waves! The only reason this description
has taken so long to mature is that
the word ``classical", in reference to the light field,
is restricted in its application to Glauber-classical
states \cite{vigier}. 
A discussion of the difference between classical and nonclassical
effects has been given in Ref.\cite{klyshko}.
The states which are produced when a nonlinear
crystal interacts with a coherent incoming beam, and, of
course, simultaneously with the vacuum, may be described
using classical Maxwell theory, but there is correlation of the outgoing
light
beams both above and below the zeropoint level.

\section{General description of parametric
down
conversion in the Wigner representation}
Type-I parametric down-conversion, in which the 
correlated signal and idler beams have the same
polarization, has been recently
studied within the framework of the Wigner function
\cite{pdc1,pdc2,pdc3}. The formalism is almost identical in the case of
type-II PDC, in which the correlated beams leaving
the nonlinear crystal are orthogonally polarized.
The process of type-II PDC can be formalized in
analogy with the
classical Hamiltonian of \cite{pdc1}
\[
H=\sum_{j=o, e}\sum_{\bf k}\hbar\omega_{j{\bf k}}
\alpha_{j{\bf k}}^{*}
\alpha_{j{\bf k}}
\]
\begin{equation}
+(i\hbar g'V
\sum_{{\bf k},{\bf k}'}
f({\bf k},{\bf k}')
exp(-i\omega_p t)
\alpha_{o{\bf k}}^{*}
\alpha_{e{\bf k}'}^{*}+ {\rm c.c.}),
\label{eqb2}
\end{equation}
$o(e)$ refers to the ordinary (extraordinary) rays.
We have taken the origin of the 
coordinate system 
at the center of the crystal, and treated the 
pump beam as an intense monochromatic 
plane wave represented by
\begin{equation}
{\bf V}({\bf r},t)=\left(V(t)exp[i({\bf k}_p
\cdot 
{\bf r}-\omega_p t)]+{\rm c.c.}\right){\bf u},
\label{eqb3}
\end{equation}
${\bf u}$ being a unit vector perpendicular to ${\bf k}_p$.
As the coherence time 
of the laser is large in comparison 
with most of the 
times involved in the process,
we shall consider 
$V(t)$ as a constant.
$g'$ is a real coupling constant defined 
so that the 
product $g'V$ has 
dimensions of frequency, and 
$f({\bf k},{\bf k}')$ is 
a dimensionless 
symmetrical function of the 
wave vectors 
inside the crystal. 
This 
function, which is related to 
the function 
$h({\bf k},{\bf k}')$
introduced in equation (8) of 
reference \cite{rubin},
is different from zero only when 
the following 
matching 
condition is 
fulfilled
\begin{equation}
{\bf k}_p \approx {\bf k}+{\bf k}'.
\label{eqb4}
\end{equation}

As is well known \cite{rubin}, there is in addition a matching
condition for frequency that is fulfilled much more rigorously,
namely,

\begin{equation}
\omega_p=\omega_{{\bf k}}+\omega_{{\bf k}'}.
\label{sarah}
\end{equation} 

On the other hand, $\alpha_{o{\bf k}}$ 
($\alpha_{e{\bf k}'}$) is the field 
amplitude for the mode with wave number
${\bf k}$ (${\bf k}'$) corresponding to
the ordinary (extraordinary) field, which is
represented as a sum of two mutually conjugate
complex {\it c}-numbers

\begin{equation}
{\bf E}_j({\bf r},t)=
{\bf E}_j^{(+)}({\bf r},t)+
{\bf E}_j^{(-)}({\bf r},t), \label{eq_w1}
\end{equation}
\begin{equation}
{\bf E}_j^{(+)}({\bf r},t)=i
\sum_{\bf k}\left(
\frac{\hbar\omega_{j{\bf k}}}{2L^3}
\right)^{\frac{1}{2}}
{\bf \epsilon}_{j{\bf k}}
\alpha_{j{\bf k}}(t)
exp(i{\bf k}\cdot{\bf r}),\,\,\,\,j=o, e,
\label{eq_w2}
\end{equation}
where $L^3$ is the normalization volume and  
${\bf \epsilon}_{j{\bf k}}$ is a polarization
vector.
Eqs. (\ref{eq_w1}) and (\ref{eq_w2}) 
correspond to 
the Heisenberg picture, where all time 
dependence goes 
in the field amplitudes
$\alpha_{j{\bf k}}^{*}(t)$ and 
$\alpha_{j{\bf k}}(t)$. For a free 
field 
this dependence has the form
\begin{equation}
\alpha_{j{\bf k}}(t)=
\alpha_{j{\bf k}}(0)
exp(-i\omega_{j{\bf k}}t), \label{eq_w3}
\end{equation}
but for interacting fields it is 
complicated and 
contains all the dynamics of the process.

The evolution of the Wigner field amplitudes
$\alpha_{j{\bf k}}(t)$ is directly given by the
Hamilton (canonical) equations of motion taking
$\sqrt{\hbar}\alpha_{j{\bf k}}(t)$ as coordinates
and $\sqrt{\hbar}\alpha_{j{\bf k}}^{*}(t)$
as canonical momenta. For instance, we get
for the extraordinary field amplitude 
$\alpha_{e{\bf k}}$:

\begin{equation}
\dot{\alpha}_{e{\bf k}}=-i\omega_{e{\bf k}}
\alpha_{e{\bf k}}
+g'V\sum_{{\bf k}'}f({\bf k},{\bf k}')
exp(-i\omega_p t)
\alpha_{o{\bf k}'}^{*},
\label{eqb5}
\end{equation}
and a similar expression holds for the 
ordinary field amplitude by 
exchanging index $e$ with $o$.

In order to calculate 
$\alpha_{e{\bf k}}(t)$ 
for all $t$ we shall take into
account that the amplitude 
$\alpha_{e{\bf k}}(t)$ 
evolves
as a free-field mode before 
entering the 
crystal and after coming out. 
We shall integrate (\ref{eqb5}) from 
$t=-\Delta t$ to $t=0$,
where $\Delta t$ is the time 
taken for the 
radiation to cross the crystal.
The initial condition is 
$\alpha_{e{\bf k}}(-\Delta t)=
\alpha_{e{\bf k}}^{{\rm (vac)}}(-\Delta t)$,
where $\alpha_{e{\bf k}}^{{\rm (vac)}}(-\Delta t)$ is
the field amplitude of the 
mode ${\bf k}$ in the incoming 
vacuum field. 

To second order in the coupling 
constant $g'$, 
that is taking the second 
term of the right side of 
(\ref{eqb5}) as a 
perturbation and retaining terms
up to order $g'^{2}$, we get 
(putting $g'\Delta t \equiv g$)
\begin{eqnarray}
\alpha_{e{\bf k}}(0)=\alpha_{e{\bf k}}^{{\rm (vac)}}(0)
+gV\sum_{{\bf k}'}f({\bf k},{\bf k}')
u[\frac{\Delta t}{2}
(\omega_p-\omega_{e{\bf k}}-
\omega_{o{\bf k}'})]
\alpha_{o{\bf k}'}^{*{\rm (vac)}}(0)
\nonumber\\
+g^2|V|^2\sum_{{\bf k}'}\sum_{{\bf k}''}
f({\bf k},{\bf k}')
f^*({\bf k}',{\bf k}'')
u[\frac{\Delta t}{2}
(\omega_{o{\bf k}'}+
\omega_{e{\bf k}''}-\omega_p)]
\nonumber\\
\times u[\frac{\Delta t}{2}(\omega_{e{\bf k}''}-
\omega_{e{\bf k}})] 
\alpha_{e{\bf k}''}^{{\rm (vac)}}(0)\,\,\,\,;\,\,\,\,g|V| \ll 1,
\label{eqb6}
\end{eqnarray}
where

\begin{equation}
u(x)=\frac{\sin{x}}{x}
exp(ix),
\label{eqb7}
\end{equation}
Equation (\ref{eqb4}) implies ${\bf k}''
\approx {\bf k}$ 
in the second order contribution 
to (\ref{eqb6}).

In the derivation of (\ref{eqb6}) 
we have taken into account that

\[
\alpha_{e{\bf k}}^{{\rm (vac)}}(0)=\alpha_{e{\bf k}}^{{\rm (vac)}}
(-\Delta t)
exp(-i\omega_{e{\bf k}}\Delta t),
\]
\begin{equation}
\alpha_{e{\bf k}}^{*{\rm (vac)}}(0)=
\alpha_{e{\bf k}}^{*{\rm (vac)}}(-\Delta t)
exp(i\omega_{e{\bf k}}\Delta t).
\label{eqb8}
\end{equation}
After $t=0$,
$\alpha_{e{\bf k}}(t)$ evolves 
as a free-field mode
\begin{equation}
\alpha_{e{\bf k}}(t)=\alpha_{e{\bf k}}(0)
exp(-i\omega_{e{\bf k}}t).
\label{eqb7a}
\end{equation}

Now, let us consider two narrow 
correlated beams called ``ordinary" and ``extraordinary",
with average frequencies 
$\omega_o$, $\omega_e$, and 
wave vectors
${\bf k}_o$, ${\bf k}_e$ respectively, 
fulfilling the matching 
conditions
\begin{equation}
\omega_o+\omega_e=\omega_p \,\,\,;\,\,\,
{\bf k}_o+{\bf k}_e={\bf k}_p.
\label{eqb12}
\end{equation}
Both light beams
contain frequencies within a range between 
$\omega_{j{\rm min}}$ 
and
$\omega_{j{\rm max}}$ ($j=o, e$), wave vectors whose
transverse components are limited
by a small upper value, 
and orthogonal polarization vectors which are 
practically independent of the wave vectors,
that is
\[
\omega_{j{\rm min}}<\omega_{j{\bf k}}
<\omega_{j{\rm max}},\;\; 
|{\bf k}^{\rm tr}|\ll 
\frac{\omega_{j{\rm min}}}{c},
\]
\begin{equation} 
{\bf \epsilon}_{e{\bf k}}\equiv{\bf \epsilon}_{e}\,\,\,\,,\,\,\,\, 
{\bf \epsilon}_{o{\bf k}}\equiv{\bf \epsilon}_{o}\,\,\,\,\,;
\,\,\,\,\, {\bf{\epsilon}}_e \cdot {\bf{\epsilon}}_{o}=0.
\label{eq_w4}
\end{equation}
We also substitute slowly 
varying
fields ${\bf F}^{(+)}({\bf r},t)$ 
(${\bf F}^{(-)}({\bf r},t)$) for the amplitudes 
${\bf E}^{(+)}({\bf r},t)$
(${\bf E}^{(-)}({\bf r},t)$), 
the relation between them 
being

\[
{\bf F}_j^{(+)}({\bf r},t)\equiv
exp(i \omega_j t){\bf E}_j^{(+)}({\bf r},t)
\]
\begin{equation}
=\left[ i \sum_{\bf [k]_{\it j}}
\left(\frac{\hbar 
\omega_{\bf k}}{2L^3}\right)^
{\frac{1}{2}} \alpha_{\bf k}(0)
exp(i{\bf k}\cdot{\bf r})
exp[i(\omega_j-\omega_{j{\bf k}})t] \right]
{\bf{\epsilon}}_j \,\,\,\,\,(j=o, e),
\label{eqc1}
\end{equation}
where $\omega_j$ is some 
appropriately chosen 
average frequency 
midway between $\omega_{\rm min}$ and 
$\omega_{\rm max}$ (see(\ref{eq_w4})).
The square brackets in 
the summation symbol 
indicates that the sum is 
restricted to the set of ${\bf k}$ 
pertaining 
to the {\it j}-beam.

It is easy to obtain 
the amplitude ${\bf F}_j^{(+)}({\bf r}_B,
t)$ in terms of the amplitude 
${\bf F}_j^{(+)}({\bf r}_A,t)$ 
at another point of the
light beam \cite{pdc1}. We find

\begin{equation}
{\bf F}_j^{(+)}({\bf r}_B,t)=
{\bf F}_j^{(+)}({\bf r}_A,t-
\frac{r_{AB}}{c})
exp(i \omega_j
\frac{r_{AB}}{c})\,\,\,\,\,(j=e, o),
\label{eqc2}
\end{equation}
where ${\bf r}_{AB}=
{\bf r}_B-{\bf r}_A$ and 
$r_{AB}=|{\bf r}_{AB}|$.

From expressions (\ref{eqb6}), 
and (\ref{eqc1}) 
we obtain
\[
{\bf F}_e^{(+)}({\bf r}, t)=F_e^{(+)}({\bf r}, t){\bf{\epsilon}}_e
\]
\begin{equation}
=\left( [1+g^2|V|^2J]F_{e}^{(+)\vac}({\bf r}, t)+
gVGF_{o}^{(-)\vac}({\bf r}, t) \right) {\bf{\epsilon}}_e,
\label{eqb13}
\end{equation}

\[
{\bf F}_o^{(+)}({\bf r}, t)=F_o^{(+)}({\bf r}, t){\bf{\epsilon}}_o
\]
\begin{equation}
\left(
[1+g^2|V|^2J]F_{o}^{(+)\vac}({\bf r}, t)
+gVGF_{e}^{(-)\vac}({\bf r}, t) \right) {\bf{\epsilon}}_o.
\label{eqb14}
\end{equation}
Here ${\bf F}_{e}^\vac$ and ${\bf F}_{o}^\vac$ are
the 
incoming vacuum fields and ${\bf F}_e$ 
(${\bf F}_o$) 
the outgoing extraordinary (ordinary) fields --- see
Fig. 1. We have
\begin{equation}
F_{e}^{(+)\vac}({\bf r},t)=
i\sum_{[{\bf k}]_e}
\left(\frac{\hbar\omega_{e{\bf k}}}
{2L^3}\right)^{\frac{1}{2}}
exp(i{\bf k}
\cdot {\bf r})exp[(\omega_e-\omega_{e{\bf k}})t]
\alpha_{e{\bf k}}^\vac(0),
\label{eqb15}
\end{equation}
and 
similarly for $F_{o}^{(+)\vac}$.
$G$ and $J$ are linear operators which are
defined as:

\begin{equation}
GF_{o}^{(-)\vac}({\bf r},t)=
i\sum_{[{\bf k}]_e}
\left(\frac{\hbar\omega_{e{\bf k}}}
{2L^3}\right)
^{\frac{1}{2}}exp(i{\bf k}
\cdot{\bf r})
exp[i(\omega_e-\omega_{e{\bf k}})t]
\beta_{\bf k},
\label{eqb16}
\end{equation}
with
\begin{equation}
\beta_{\bf k}=
\sum_{[{\bf k}']_o}
f({\bf k},{\bf k}')
u[\frac{\Delta t}{2}(\omega_p-
\omega_{e{\bf k}}-
\omega_{o{\bf k}'})]
\alpha_{o{\bf k}'}^{*\vac}(0),
\end{equation}
and
\begin{equation}
JF_{e}^{(+)\vac}({\bf r},t)=
i\sum_{[{\bf k}]_e}
\left(\frac{\hbar\omega_{e{\bf k}}}
{2L^3}\right)^{\frac{1}{2}}
exp(i{\bf k}\cdot{\bf r})
exp[i(\omega_e-\omega_{e{\bf k}})t]
\gamma_{\bf k},
\label{eqb17}
\end{equation}
with
\[
\gamma_{\bf k}=
\sum_{[{\bf k}']_o}\sum_{[{\bf k}'']_e}
f({\bf k},{\bf k}')
f^*({\bf k}',{\bf k}'')
u[\frac{\Delta t}{2}(\omega_{o{\bf k}'}+
\omega_{e{\bf k}''}-\omega_p)]\times
\]
\begin{equation}
\times u[\frac{\Delta t}{2}(\omega_{e{\bf k}''}-
\omega_{e{\bf k}})]
\alpha_{e{\bf k}''}^\vac(0).
\label{eqbb17}
\end{equation}

From (\ref{eqb13}) we see that the 
outgoing 
extraordinary beam, to order $g^2$, 
consists of three parts: {\em i}) 
A zeropoint 
radiation with 
amplitude $F_{e}^{(+)\vac}$, which
passes through the crystal without any 
change; {\em ii}) a radiation produced by
the nonlinear interaction 
(mediated by the 
crystal) between the laser beam,
with amplitude $V$, and the zeropoint 
radiation,
with amplitude $F_{o}^{(-)\vac}$,
entering the 
crystal in the direction of the
ordinary beam; {\em iii}) one part 
which just 
modifies a little (to order $g^2$)
the amplitude $F_{e}^{(+)\vac}$. The
ordinary beam 
is constituted in a similar
manner.

Now, let us consider 
the correlation properties of the
fields, which are identical to the ones
calculated in
a previous work \cite{pdc2}:

\begin{center}
a) Autocorrelations.
\end{center}

Taking the extraordinary
field 
${\bf F}_e^{(+)}({\bf r}, t)=
F_e^{(+)}({\bf r}, t){\bf{\epsilon}}_e$ 
at a point ${\bf r}$
and times $t$ and $t'$, we have:
\[
\langle F_e^{(+)}({\bf r},t)
F_e^{(-)}({\bf r},t') \rangle 
-\langle F_{e}^{(+)\vac}({\bf r},t)
F_{e}^{(-)\vac}({\bf r},t') \rangle =
\]
\[
2g^2|V|^2\langle GF_{o}^{(-)\vac}({\bf r},t)
G^*F_{o}^{(+)\vac}({\bf r},t') \rangle
\equiv g^2|V|^2\mu_e(t'-t),
\]
\begin{equation}
\langle F_e^{(+)}({\bf r},t)
F_e^{(+)}({\bf r},t') \rangle=0.
\label{eqb}
\end{equation}
Here ``$\langle \rangle$" means an average using the
Wigner function in the vacuum state as probability density.
$\mu_e(t-t')$ is a correlation
function 
which goes to zero when $|t'-t|$ is greater than the
correlation 
time of the extraordinary beam, $\tau_e$. Similar expressions
hold for the ordinary field by exchanging the indices
``$e$" and ``$o$".

\begin{center}
b) Crosscorrelations.
\end{center}

Taking the extraordinary 
(${\bf F}_e^{(+)}({\bf r}, t)=
F_e^{(+)}({\bf r}, t){\bf{\epsilon}}_e$)
and ordinary 
(${\bf F}_o^{(+)}({\bf r}, t)=
F_o^{(+)}({\bf r}, t){\bf{\epsilon}}_o$)
fields at the center
of the crystal ${\bf r}={\bf r}'={\bf 0}$ and
times $t$ and $t'$, we have:
 
\[
\langle F_e^{(+)}({\bf 0},t)
F_o^{(+)}({\bf 0},t') \rangle 
=gV\nu(t'-t).
\]
\begin{equation}
\langle F_e^{(+)}({\bf 0},t)
F_o^{(-)}({\bf 0},t') \rangle=
\langle F_e^{(-)}({\bf 0},t)
F_o^{(+)}({\bf 0},t') \rangle=0.
\label{eqc}
\end{equation}
Here $\nu(t'-t)$ is a function
which vanishes when $|t'-t|$ is greater
than the coherence time between the extraordinary
and ordinary beams. From (\ref{eqc}) it is possible
to derive all crosscorrelations at different
points ${\bf r} \neq {\bf r}'$ by using (\ref{eqc2}).

Finally, the quantum theory of detection 
in the Wigner representation gives us the following
results for single and joint detection 
probabilities:

\begin{center}
a) Single probability.
\end{center}

The following result
is a general expression
for calculating single probabilities per unit time 
in the Wigner
representation:

\begin{equation}
P_1({\bf r}_1,t) \propto
\langle I({\bf r}_1,t)-I_0({\bf r}_1)\rangle,
\label{eqd}
\end{equation}
where $I({\bf r}_1,t)=|{\bf E}^{(+)}({\bf r}_1,t)|^2=
|{\bf F}^{(+)}({\bf r}_1,t)|^2$, and 
$I_0({\bf r}_1)$ is the intensity of the vacuum
field at the position of the detector. 

\begin{center}
b) Joint probability.
\end{center}

It can be proved that
in PDC experiments
\begin{equation}
P_{12}({\bf r}_1,t;{\bf r}_2,t+\tau) \propto
\langle\{I({\bf r}_1,t)-I_0({\bf r}_1)\}
\{I({\bf r}_2,t+\tau)-I_0({\bf r}_2)\}\rangle.
\label{eqd1}
\end{equation}
By taking into account that the Wigner 
fields amplitudes are Gaussian processes, and
neglecting fourth
order terms in $g$, we have \cite{pdc3}:

\begin{equation}
P_{ab}({\bf r}_1,t;{\bf r}_2,t+\tau)
\propto \sum_{\lambda}\sum_{\lambda'}
|\langle F_{\lambda}^{(+)}({\bf r}_1,t)
F_{\lambda'}^{(+)}({\bf r}_2,t+\tau)\rangle|^2,
\label{eq14}
\end{equation}
where $\lambda$ and $\lambda'$ are 
polarization indices.
\section{Tests of Bell's inequalities using polarization correlation}
Most experiments to test Bell's inequalities with nonlinear crystals
performed hitherto
have used type-I parametric down-conversion in which
the two correlated beams have the same polarization. In
\cite{pdc1,pdc2,pdc3}, experiments of this kind were analyzed
in the Wigner function formalism. However,
more recent experiments, using type-II phase matching,
provide a more direct way to generate ``entangled-photon" states.
Type-II experiments are themselves of two types. In the first,
that is collinear type-II PDC, the crystal is
oriented so that the ordinary and extraordinary radiation cones are
mutually tangent in the direction of the pumping beam. To date, nearly all
type-II experiments have used collinear phase matching
\cite{shih}. On the other hand \cite{kwiat},
in noncollinear type-II phase matching, the two
cones intersect along two
directions, and this gives rise to an entangled state in the
polarization (see Fig.2). It has been claimed that such a source
produces true entangled states, capable of violating Bell's inequalities.

The experimental outline is shown in Fig.3. The two beams
``$1$" and ``$2$", in which the ordinary and extraordinary cones
intersect, are selected and sent to two polarizers $P_1$ and $P_2$
oriented at angles $\phi_1$ and $\phi_2$ with respect to
the polarization of the extraordinary ray. Coincidence rates
were measured as functions of angles $\phi_1$ and $\phi_2$. 
In \cite{kwiat} additional optical devices, that is half- and
quarter-wave plates, were used in order to produce
four different Bell
states, but we shall confine our analysis to just one of
these states, namely the one which uses no additional devices.

Let us see how the entangled state is
represented in the Wigner formalism. The two beams, coming out of the
crystal along the directions where the ordinary and 
extraordinary cones intersect, are given by

\[
{\bf F}_1^{(+)}({\bf 0},t)=F_e^{(+)}({\bf 0},t)
{\bf i}+F_{o'}^{(+)}({\bf 0},t){\bf j},
\]
\begin{equation}
{\bf F}_2^{(+)}({\bf 0},t)=F_{e'}^{(+)}({\bf 0},t)
{\bf i'}+F_{o}^{(+)}({\bf 0},t){\bf j}',
\label{no}
\end{equation}
where ${\bf i,i'}$ represent the polarizations of the extraordinary
beams and ${\bf j}$, ${\bf j}'$ the polarizations
of the ordinary beams. The essential point is that the extraordinary
component, $F_e^{(+)}$, of the first ray and the ordinary component, $F_o^
{(+)}$,
of the second ray are conjugated, and therefore correlated. Similarly,
$F_{o'}^{(+)}$ and $F_{e'}^{(+)}$ are correlated, but $F_e^{(+)}$ ($F_o^{(
+)}$)
is uncorrelated to $F_{e'}^{(+)}$ ($F_{o'}^{(+)}$). 

When a polarizer oriented at angle $\phi_1$ 
to the horizontal is placed
in front of the detector $D_1$,
the field at $D_1$ (placed at ${\bf r}_1$)
at time $t$ is 

\[
{\bf F}^{(+)}({\bf r}_1,t)=[{\bf F}_1^{(+)}({\bf r}_1, t)\cdot
(\cos{\phi_1}{\bf i}+\sin{\phi_1}{\bf j})]
(\cos{\phi_1}{\bf i}+\sin{\phi_1}{\bf j})
\]
\[
=exp[i\omega(d/c)][F_e^{(+)}({\bf 0},t-\frac{d}{c})\cos{\phi_1}+
F_{o'}^{(+)}({\bf 0},t-\frac{d}{c})\sin{\phi_1}]
\]
\begin{equation}
(\cos{\phi_1}{\bf i}+\sin{\phi_1}{\bf j})
\label{nor}
\end{equation}
Here we recall that the action of a polarizer is to project
the electric field vector on the polarization direction

In the same way,
we write for the
field at the detector $D_2$ (placed at ${\bf r}_2$)
at time $t+\tau$

\[
{\bf F}^{(+)}({\bf r}_2,t+\tau)=[{\bf F}_2^{(+)}({\bf r}_2, t+\tau)\cdot
(\cos{\phi_2}{\bf i'}+\sin{\phi_2}{\bf j}')]
(\cos{\phi_2}{\bf i'}+\sin{\phi_2}{\bf j}')
\]
\[
=exp[i\omega(d/c)]
[F_{e'}^{(+)}({\bf 0}, t+\tau-\frac{d}{c})\cos{\phi_2}+ 
F_{o}^{(+)}({\bf 0}, t+\tau-\frac{d}{c})\sin{\phi_2}]
\]
\begin{equation}
(\cos{\phi_2}{\bf i'}+\sin{\phi_2}{\bf j}')
\label{norr}
\end{equation}

In order to calculate the joint probability
we combine Eqs. (\ref{eq14}), (\ref{nor}), and (\ref{norr}), and
take into account the correlation properties of the fields
given by (\ref{eqb}) and (\ref{eqc}).
After some easy algebra 
and an integration of
$P_{12}(\tau)$ over the detection window,
we obtain the
coincidence probability

\begin{equation}
P_{12}=K\sin^2({\phi_2}+{\phi_1}),
\label{norrr}
\end{equation}
$K$ being a constant. This
expression is similar to the one obtained
in \cite{kwiat}, and corresponds to $100\%$ contrast. This type
of correlation is usually claimed to violate a Bell inequality 
(but see the discussion section). In the actual experiment  
a violation of the inequality by $100$ standard deviations
is reported.
\section{The detection problem in the Wigner representation}
The detection probability in quantum optics is
usually written in terms of the normally ordered expression

\begin{equation}
P \propto \int_{0}^{\Delta t} \langle \hat{E}^{(-)}(t) \hat{E}^{(+)}(t) 
\rangle dt,
\label{eq1}
\end{equation}
where $\hat{E}^{(+)}(t)$ is the Heisenberg operator of the electric field
at the detector, $\hat{E}^{(-)}(t)$ its hermitian conjugate and $\Delta t$
is the detection time window. For simplicity we consider a point detector
and, therefore, ignore the standard volume integral in (\ref{eq1}).

When we pass to the Wigner representation, the normally ordered 
expression should be written in terms of a symmetrically ordered 
expression minus a commutator. Then we may replace (see Eq.(\ref{eqd}))
the
Heisenberg operators by random wave amplitudes and, after some rather
trivial algebra, we get

\begin{equation}
P \propto \int_{0}^{\Delta t} \langle I(t)-I_0 \rangle dt,
\label{eq2}
\end{equation}
where $I(t)=|E(t)|^2$ is the intensity of the field arriving at the
detector and $I_0$ is a constant corresponding to the average intensity
of the zeropoint, i.e. the intensity that would arrive at the
detector if 
all
light sources, such as lasers, were switched off.

If the Wigner function is positive definite, $I(t)$ may be interpreted
as a stochastic process, which makes possible a wavelike interpretation
of the propagation of light. This is the case in all experiments involving
parametric down conversion. However, there remains a
problem for a wavelike interpretation of the detection process,
because $I(t)-I_0$ is not positive-definite. This means we cannot assume
that $I(t)-I_0$ is proportional to a detection probability. Nevertheless,
it is easy to show that the average $\langle I(t)-I_0 \rangle$ is
nonnegative-definite. Now we consider the usual case where $I(t)$ is a 
stationary stochastic process. It then follows (see Ref.\cite{mandel},
page 49)
that it is ergodic. We could consider
substituting time averages for
ensemble averages, thereby obtaining

\begin{equation}
P \propto  \langle I(t)-I_0 \rangle =\lim_{T\rightarrow \infty}
\frac{1}{T} \int_{0}^{T} [I(t)-I_0]dt.
\label{eq3}
\end{equation}
This shows that, for every member of the ensemble of stochastic wave 
amplitudes (except for a subensemble of zero measure), the
time-averaged photodetection probability precisely equals the
ensemble averaged one $\langle I(t)-I_0 \rangle$.

In practice we do not have a time-average but something which is
almost equivalent, namely an integration over the detection window
$\Delta t$. A typical detection window lasts more than one nanosecond,
so that the dimensionless quantity $\omega\Delta t $, where $\omega$ is
the
frequency of visible light, is of the order of $10^7$. It is true that
taking the limit $T\rightarrow \infty$ is not the same as taking 
$T \approx 10^7$ (in dimensionless units), but the 
difference must be small. Since the right hand
side of (\ref{eq3}) is nonnegative-definite, this means
that the finite-time average

\begin{equation}
\frac{1}{\Delta t} \int_{0}^{\Delta t} [I(t)-I_0]dt
\label{eq4}
\end{equation}
takes negative values only with a very small probability.
Now let us modify the standard detection probability of Eq.(\ref{eq2})
so that
\begin{equation}
P \propto\left\{ \int_{0}^{\Delta t} [I(t)-I_0]dt\right\}_+ ,
\label{eq5}
\end{equation}
where the notation $\{\}_+$ indicates that we replace
the contents of the brackets by zero if their value is negative.
Then (\ref{eq5})
will predict rather more photocounts than the
standard quantum detection theory,
but it seems not unreasonable to assume that
these additional counts correspond to a part of the dark
rate at the
detector. As a matter of fact, 
the quantum theory of detection, leading to (\ref{eq1}), 
involves first order perturbation theory. Therefore we may assume that
quantum theory also predicts some dark
background in photocounters when higher-order processes are taken into
account, because the detector may be activated by vacuum
fluctuations. A detailed study of such a background would
have to enter fully into the electronic band structure of the
photodetector material\cite{rieke}; a treatment based on the 
photoelectric effect for single atoms is clearly inadequate.

\section{Discussion}
In the preceding sections we have outlined a theory of both
the propagation and detection of light which is consistently
local realist in the sense defined, for example, by Clauser
and Shimony\cite{clauser}. 
In fact, eq.(\ref{eqd1}) has the standard form introduced
by Bell in his definition of local hidden variables models, in particular

\begin{equation}
P_{12}=\langle P_1(\lambda, \phi_1)P_2(\lambda, \phi_2) \rangle=
\int \rho(\lambda)P_1(\lambda, \phi_1)P_2(\lambda, \phi_2)d\lambda,
\end{equation}
where the amplitudes $\alpha_{j, {\bf k}}$ involved in 
$I({\bf r}, t)$, via $E({\bf r}, t)$ (see eq.(\ref{eq_w2})
and the line after eq.(\ref{eqd})) play the role of the hidden variables
$\lambda$. On the other hand, our theory is also
in almost perfect
one-to-one correspondence with the standard Hilbert-space
theory, the only difference being the modification in the
detection probability which we have proposed in Eq.(\ref{eq5}).
Apart from a small dark rate, which is, in
any case, a feature of all real experimental situations,
this theory gives singles and coincidence rates agreeing
with the standard theory . 
Therefore we have here an apparent contradiction. On the one hand,
a Bell inequality is violated
according to Ref.\cite{kwiat}, which implies that no local hidden 
variables model exists for the experiment. On the other hand we have 
an explicit local hidden variables model, namely the quantum model
in the Wigner representation with the modification of (\ref{eq5}).
What is the resolution of this apparent contradiction? The reason is
that the Bell inequality which is actually violated in the experiment is
not a genuine Bell inequality derived from the assumptions of realism
and locality alone (like inequality (4) of Clauser and Horne
\cite{clauhorn}), but a homogeneous inequality involving additional
assumptions (like inequality (11) of Clauser and Horne). One of these
assumptions, rather than local realism, is what is violated in our
LHV model.  

The present authors have been insisting
for many years now \cite{marshsant} that these auxiliary
assumptions are not only unreasonably restrictive but also
incorrect. We believe that the results reported in the present article
vindicate our point of view.

As is well known a genuine Bell inequality cannot be tested in
experiments with visible light, due to the low efficiency of the
detectors presently available. The conventional wisdom is that this is
just
a technical problem that will be solved in the near future. However, the
existence of an LHV model for the quoted experiment \cite{kwiat} which 
does not rest upon the low efficiency of the detectors, but on the
existence of some unavoidable amount of dark rate (see Eq.(\ref{eq5}))
shows that any future reliable test of LHV theories should involve 
detectors having both high efficiency and low dark rate.

Now, we turn to the interpretation of entanglement in the
Wigner representation. We saw (see Eqs.(\ref{eqc}) and (\ref{no})) that
``photon entanglement" is just correlation between two light beams.
Then,
what is
the difference between ``classical" correlation and entanglement? In
quantum
optics what is usually called ``classical" is light having a (Glauber) 
P-representation which is positive definite, and ``classical correlation"
usually means a correlation between the P-distribution functions
of the two light beams. Here we see that ``local realist theories" is a
class
much bigger than standard ``classical theories". In particular,
the quantum model following from the Wigner representation interpreted
as a (positive) probability distribution  allows for correlations
much stronger than ``classical correlations". As in our functions
$F_e^{(+)}$ and $F_o^{(+)}$ of Eq.(\ref{eqc}) they involve
correlation of 
the
zeropoint part of the electromagnetic field, and this is ``entanglement".

Of course, the obvious objection to these arguments is that they
cannot be
extended
to cases where the Wigner function is negative. Elsewhere we have
conjectured
that this never happens in actual experiments. We refer to our publication
\cite{vigier} for details. 

\vspace{0.5cm}

{\bf Acknowledgement}

We acknowledge financial support of CAICYT 
Project No. PB-95-0594 (Spain).

\newpage
{\Large
{\bf Figure captions}}
     
\begin{itemize}
\item[] Fig. 1. The process of parametric down conversion.

\item[] Fig. 2. Polarization entanglement in noncollinear
type-II down conversion.

\item[] Fig. 3. Tests of Bell's inequalities using noncollinear
type-II down conversion.

\end{itemize}

Figures have not yet been included, as the authors have to
master the appropriate technology first. We hope it will
not take very long!

\end{document}